\RequirePackage[dvipsnames,table,prologue]{xcolor}
\documentclass[runningheads]{llncs}
\usepackage{amsmath}
\usepackage{graphicx}
\graphicspath{{graphics/}}
\usepackage[inline]{enumitem}
\usepackage{fancyvrb}
\usepackage{microtype}
\usepackage{wrapfig}
\usepackage[
sortcites,
backend=biber,
style=numeric,
firstinits=true,
url=false,
isbn=false,
safeinputenc,
]{biblatex}

\newcommand{\keywordSeparator}{, }

\newcommand{\fullTitle}{%
10 Years Later: The Mathematics Subject Classification and Linked Open Data%
}
\ifdefined\titlerunning
  \titlerunning{10 Years Later: MSC \& LOD}
\fi

\makeatletter
\@ifclassloaded{acmart}{%
  \title[\shortTitle]{\fullTitle}%
}{%
  \title{\fullTitle}%
  \renewcommand{\keywordSeparator}{\and}
}
\makeatother

\begin{document}

\makeatletter
\newcommand{\printfnsymbol}[1]{%
  \textsuperscript{\@fnsymbol{#1}}%
}
\makeatother
\author{Susanne Arndt\thanks{All authors contributed equally.}\inst{1}
\and
Patrick Ion\printfnsymbol 1 
\and
Mila Runnwerth\printfnsymbol 1 \inst{1} \and
Moritz Schubotz\printfnsymbol 1 
\and
Olaf Teschke\printfnsymbol 1 \inst{3}}
\authorrunning{S. Arndt et al.}
%
\institute{Princeton University, Princeton NJ 08544, USA \and
Springer Heidelberg, Tiergartenstr. 17, 69121 Heidelberg, Germany
\email{lncs@springer.com}\\
\url{http://www.springer.com/gp/computer-science/lncs} \and
ABC Institute, Rupert-Karls-University Heidelberg, Heidelberg, Germany\\
\email{\{abc,lncs\}@uni-heidelberg.de}}

\institute{TIB Leibniz Information Centre for Science \& Technology, Hanover, Germany\\
\email{\{first.last\}@tib.eu}\\ \and
University of Michigan, Michigan, USA \and
zbMATH, FIZ Karlsruhe, Karlsruhe, Germany \\
\email{\{first.last\}@fiz-karlsruhe.de}}

\maketitle
\begin{abstract}
Ten years ago, the Mathematics Subject Classification  MSC 2010 was
released, and a corresponding machine-readable Linked Open Data collection was published using the Simple Knowledge Organization System (SKOS).
Now, the new MSC 2020 is out.
 
This paper recaps the last ten years of working on machine-readable MSC data and presents the new machine-readable MSC 2020.
We describe the processing required to convert the version of record, as
agreed by the editors of zbMATH and Mathematical Reviews, into the Linked Open Data form we call MSC2020-SKOS.
The new form includes explicit marking of the changes from 2010 to 2020, some translations of English code descriptions into Chinese, Italian,
and Russian, and extra material relating MSC to other mathematics classification efforts.
We also outline future potential uses for MSC2020-SKOS in semantic indexing and sketch its embedding in a larger vision of scientific research data.


\end{abstract}

\section{Introduction}\label{sec:intro}

The Mathematics Subject Classification (MSC) is a subject-specific indexing schema for mathematics.  Like universal library classifications such as the Dewey Decimal Classification\footnote{\url{https://www.loc.gov/aba/dewey/}}, the MSC can be used to assign to mathematical knowledge, whether in a printed book, electronic journal article, or conference recording, codes representing topics (categories or classes of mathematical items) covered within the discipline of mathematics or closely related research areas.
The MSC is well established in the community and used by scientists, publishers, and librarians. Beyond indexing mathematical research resources, it is also employed to describe specialties desired for academic positions or content of conference talks. It plays a very useful role in matching papers to suitable reviewers, especially at the two major post-publication reviewing services in mathematics who are responsible for the MSC. 

In 2020, the fourth official major release was published by the executive editors of Mathematical Reviews (MR) and zbMATH \cite{dunnehulek2020}. Although minor modifications are implemented as needed by MR and zbMATH, major revisions are released each decade. The editorial process is governed by MR and zbMATH in collaboration. Suggestions from mathematicians or knowledge engineers are submitted, both by mail and at the \url{msc2020.org} website, and their adoption is discussed subsequently.  As of Friday, 7 May 2021, all open issues resulting from the feedback on the MSC2020 revision have been resolved by MR and zbMATH, and MSC 2020 has its final form at last;  a definitive SKOS form can now be made.

The MSC is organised into three hierarchical levels: The 63 top levels list all major mathematical fields as topics.  They range from the foundations of mathematics to algebra, analysis, geometry and topology, and also include a wide array of topics concerning mathematics in its applications.  The MSC is fundamentally a simple three-level tree;
it can be thought of as rooted in a node for all mathematics of which the top-level classes are children, and is actually a rooted labelled planar tree in mathematical terms.
The 1.037 second-level classes represent sub-fields of each speciality.
The 5.503 third-level classes reflect the intricacies of sub-fields, for instance, 
the subtleties of different views on real or complex functions.   In addition there are 
cross-references from one topic to another of various types.

Each class is assigned a code in a notation with five characters, e.g., 68-XX or 03B25.
The first two digits indicate the top level classes numbered from 00 to 97 with gaps that leave room for future developments; the remaining three characters of top level codes are `-XX'.  For instance classes 19, 37 and 74 have all been added since 1980.
In practice, MR and zbMATH editors, and others, often omit these last three characters, e.g., one uses `68' as a short form for `68-XX' to refer to Computer Science, or 11 for Number Theory.(categories or classes) 

The second level classes are of two kinds (we are using the digits 99 as placeholders to. illustrate the formats): 
\begin{description}
\item [99-99] Second-level classes with codes that begin with the two digits of a top-level class, then a hyphen ``-", and are followed by two digits; these are used for formal meta descriptions providing special categories for such items as textbooks, historical works, e.~g., 11-03 for history of number theory or  11-06 for conference proceedings in number theory.  MSC 2020 extended these facet classes in accordance with the needs of the community, and standardized their relation with subject classes, by the introduction of further numerical codes, e.~g,  '-10' classes for mathematical modeling and simulation, and '-11' for  research data across the top level classes (e.~g., 11-11 for research data in number theory).
\item [99Axx] Classes whose codes have the structure of two digits, an uppercase letter, followed by two lowercase x's; these indicate specific mathematical areas within a top-level class.
    For example, 11Axx is the second-level category for `Elementary number theory'.
\end{description}
\begin{wrapfigure}{r}{0.5\textwidth}
\centering
\includegraphics[width=.5\textwidth]{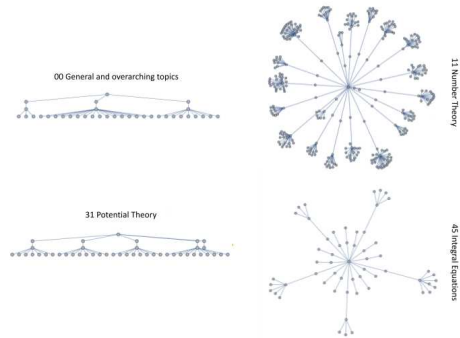}
\caption{Example visualisations of the hierarchical distribution within four MSC (2010) classes provided by \cite{Schreiber2011}.} \label{fig:MSCvis}
\end{wrapfigure}
Finally, the third-level classes are the narrowest and most specialised categories for annotating mathematical information. Their codes can be recognised from an uppercase letter in third position followed by two decimal digits, e.~g., 03B25 for ``Decidability of theories and sets of sentences [See also 11U05, 12L05, 20F10]''.

Figure~\ref{fig:MSCvis} shows the inner hierarchical structure of four selected MSC classes 00, 11, 31 and 45;  one sees the subtrees rooted on the major classes 
displayed, and that Number Theory has been much more finely coded than the 
other three subjects.\footnote{That this is so reflects that the AMS published three
ever larger collections of Reviews in Number Theory, edited by William J. Leveque and Richard K.Guy.  As these thousands of reviews were collected together it became
possible to distinguish nuances in sub-sub-topics, and an extension of the codes 
beyond 5 characters was suggested; that is the 3-level tree was to be extended
with more levels.}

While no top-level class has been changed in the MSC2020 version, several second-level classes have been added and reorganized, as have many third-level classes. One may note the large variation in the granularity of the classifications provided, which reflects the different needs of the respective mathematical communities. Accordingly, the number of documents assigned particular third-level classes varies a lot and is influenced by the different publication cultures within mathematics. Hence, one must be very careful when performing quantitative scientometric analysis using MSC classes -- pure comparison of numbers will often be very misleading.

Apart from the reorganization of the second-level facet classes mentioned above, another significant feature of MSC2020 has been a complete overhaul of the descriptive texts with the aim of more precise disambiguations. The experience from the SKOSification of MSC2010 has been extremely helpful in this regard. Now, every single description of a 
third-level class is unique regardless of its top level or second level components; this used not to be so as descriptive text was reused. The relations between MSC classes have been standardized accordingly.

Class name changes are subject to careful editorial review by zbMATH and MR and not made lightly.  The problem of shifting meaning, and community attention, within mathematics is one of those that management of this branch of knowledge has to deal with.

\subsection{Linked Open Data}
Abstractly seen, Linked Open Data (LOD) is a structured framework for  providing freely accessible, machine interpretable, and inter-operable data.  Firstly, LOD data is published under an explicit open license. Then, every concept used in describing the data has a Unique Resource Identifier (URI), and lastly, the data are encoded according to Semantic Web standards like the Resource Description Framework (RDF) or the Web Ontology Language (OWL).

In our work, adopting the principles of LOD, we worked across the layers of the frameworks RDF and its formalisation, the Simple Knowledge Organization System (SKOS)\footnote{For the SKOS specification see \url{https://www.w3.org/TR/skos-reference/}, and for the `rules of engagement' consult \url{https://www.w3.org/TR/2009/NOTE-skos-primer-20090818/}.}, until we arrived at a concrete implementation in the form of a Turtle serialisation. Turtle can be used to refer both to the syntax used to describe RDF data (schemes) and the file format.

\subsection{MSC 2010 SKOS model}
The first conversion of the MSC to RDF Linked Data was published in 2012 as a Turtle serialisation \cite{MCS2010LOD, LangeIDBSKA12}.
The motivation was to encourage reuse, maintenance, and versatile access with a low-threshold according to the best practices of the time \cite{MCS2010LOD}.
The modeling decisions made then have been comprehensively documented in \cite{LangeIDBSKA12}.
SKOS Core was chosen as a point of departure and gradually extended to represent the MSC's semantic subtleties.
This resulted in an extension \emph{mscvocab} defining inter alia: \emph{related part of}, \emph{see also}, and \emph{see mainly}.
Another subject specific characteristic is the use of the Mathematical Markup Language (MathML)\footnote{\url{https://www.w3.org/TR/MathML3/}} within a SKOS vocabulary.

The  data required to represent the complete model for MSC 2010 are publicly available\footnote{\url{http://msc2010.org/resources/MSC/2010/info/}}.

This approach inspired other projects tackling semantification of mathematics according to LOD principles, for example \emph{OntoMathPRO} \cite{Nevzorova2014} or \emph{coli-conc} \cite{Balakrishnan_2018}.  The \emph{Encyclopedia of mathematics} used to annotate its records with the SKOSified MSC 2010 via SPARQL query \cite{Rehmann2016}.

\section{MSC 2020 SKOSification}\label{sec:method}
The latest release gives us an opportunity to revisit and improve the Linked Open Data form of the MSC.
We challenge, and largely, confirm the concept modeling decisions made before.  Moreover, we add state-of-the-art metadata to describe the different SKOS versions of each release, and license information to legally describe and verify its open access use.
Finally, we justify our approach with specific use cases which rely on an RDF Linked Data representation of the MSC 2020 (including back-links to its history).

 The first SKOS model showed a level of sophistication that we would like to adhere to, i.~e. there will be no unnecessary modifications. 

An obvious reason for a new SKOS version are the MSC modifications made between the two releases MSC 2010 and 2020.
In addition, particularly for further reuse in German speaking countries, we are adding German labels. 

We challenge, and largely, confirm the concept modeling decisions made before.  Moreover, we add state-of-the-art metadata to describe the different SKOS versions of each release, and license information to legally describe and verify its open access use.

Significant reasons for a revised SKOS formalisation of the MSC are three specific use cases in libraries: 
\begin{enumerate}
    \item Automated subject indexing of mathematical library inventories with the toolkit annif \cite{Suominen_2019}. The optimal input format for classifications and vocabularies is a Turtle serialisation.
    \item Providing a SKOS version compatible with the MSC2010's for the extensive mapping project \emph{coli-conc}, including its mapping editor \emph{Cocoda} \cite{Balakrishnan_2018}. The project already records the Dewey Decimal Classification, the MSC 2010, and Wikidata and includes several mappings between classifications. 
    \item The \emph{Open Research Knowledge Graph} (ORKG) aims at providing machine interpretable semantic graphs for research questions and individual papers in order to make them comparable using standardised queries \cite{Jaradeh_2019}. The quality of such a graph depends on the authority files or thesauri upon which it is built. Since the ORKG follows the LOD principles a SKOS formalisation of the MSC would be compatible and could be applied to graphs derived from mathematical scholarly knowledge.   
\end{enumerate}

However, those reasons do not touch upon the structural requirements for a sustainable MSC 2020 in the Semantic Web.
In a first step, we tidied up minor bugs, e.~g. spaces in URIs to guarantee a valid and consistent serialisation.
Then, we made well-founded conceptual adjustments.  One long-term goal is to reduce the effort of moving from one MSC release to another.

\subsection{Tools for implementation \& our procedure}
We provide a \emph{GitHub}\footnote{\url{https://github.com/runnwerth/MSC2020_SKOS}} repository containing the Turtle file itself and its extensions, but also with the appropriate automation scripts.
Furthermore, the process is documented in some wiki pages.
We chose the widely used platform GitHub, because -- aside from being a suitable repository and archive -- its issue tracking has been shown to support a transparent and inclusive editorial process of vocabulary management \cite{Frey_2018}.
implementation
Different tools were used to create the SKOS form than those used for MSC 2010.
The Google big data handler OpenRefine was used to massage Excel files this time,
whereas Perl and Python scripts acting on a \TeX\ source file for print were used 
for the first SKOS version.   I could be argued that OpeRefine allows you to approach 
the task with a different skill set than that of a developer/ IT specialist.  One needs to pay less and it can be claimed to consume fewer person hours to get the job done.

In principle, this could have led to notable differences in the complexity of the 
process, and the evolution of LOD techniques might have changed what could
be done too.  It turned out that both processes led to similar results.  Re-examination 
of the data differently did provide additional quality control and a number of small
errors were caught this way.

zbMATH provided database excerpts of MSC 2010, MSC 2020, and their respective modifications as \emph{Excel} files in three spreadsheets: The two spreadsheets documenting the MCSs, respectively, contain a column for the five-character MSC code, a column for the code's descriptive text, and an annotation column, with e.~g. references as given in the MSCs. The third spreadsheet traces the changes between MSC releases by class, and splits them into five categories
\begin{enumerate}
    \item splitting a class into at least one other class,
    \item a newly introduced class,
    \item a moved class,
    \item merged classes, and
    \item deleted classes.
\end{enumerate}

To begin we prepared a Turtle file with nothing but a preamble declaring namespaces, imports, and metadata using the ontology editor Prot{\'{e}}g{\'{e}}\footnote{\url{https://protege.stanford.edu/}} \cite{Musen15}. 

Then, we used \emph{OpenRefine} to create all the basic triples for the MSC 2020 \cite{ham2013openrefine, cory2013}: identifiers for every class, \emph{rdf:type} of elements, further elementary triples such as \emph{skos:prefLabel, skos:notation, skos:scopeNote, skos:inScheme}. The transformation process, expressed in the \emph{General Refine Expression Language (GREL)}, is minutely tracked and can be exported in JSON for reuse. The output is a table in which columns contain valid RDF triples. These can be imported into our prepared Turtle file. Alternatively, an RDF extension for OpenRefine can be used. It allows for a mapping from table data to imported RDF vocabularies (e.g. SKOS). Mapped data can then be exported as a Turtle file\footnote{\url{https://github.com/stkenny/grefine-rdf-extension}}.

The third step was modeling the different categories of internal references between MSC classes, i.~e. the cross-references \emph{mscvocab:seeMainly, mscvocab:seeAlso, mscvocab:seeFor}. In the first SKOS version the master authority was a \TeX{} file whereas, here, we derived the internal references from the \emph{description} or \emph{annotation} column of the MSC's databases provided to us. Table~\ref{tab:03B45} shows the row for the MSC class 03B45. 

Textual information in the \emph{description} column was mostly homogeneous and standardised to a certain degree and could therefore easily be transformed into triples working with \emph{mscvocabs} object properties. However, there were also irregularities, e.~g., there seems to be no standardised order of information types in the column, and on some occasions, there were deviations from the regular notation.

\begin{table}[]
    \centering
    \begin{tabular}{|p{1cm}|p{4cm}|p{7cm}|}
    \hline
        \textbf{code} & \textbf{text} & \textbf{description} \\ \hline
        03B45 & Modal logic (including the logic of norms) & Modal logic (including the logic of norms) \{For knowledge and belief, see 03B42; for temporal logic, see 03B44; for provability logic, see also 03F45\} \\ \hline
    \end{tabular}
    \caption{Original data for the MSC class 03B45. The \emph{description} column is broken down in several information categories that are modeled individually.}
    \label{tab:03B45}
\end{table}

To address the `multi-valuedness' and irregularities in this column, we split the cells according to shared regularities until there was only an irregular remainder left. The values in these new columns can then be used as objects of triples, and are either URIs or literals. 

The most complex references were qualified ones. In the example  given in Table~\ref{tab:03B45} the reference to the class 03B44 is restricted to `temporal logic'. To account for this limited scope of a reference we adopted the MSC 2010's property \emph{mscvocab:seeConditionally}:

[fontsize=\small]
\begin{verbatim} 
<http://msc.org/resources/MSC/msc2020/03B45> 
    mscvocab:seeConditionally <http://msc.org/resources/MSC/msc2020/03B42> .
\end{verbatim}

While this is straightforward, providing the condition requires an additional structural solution, e.~g., an annotation on a relation or referring to the relation by its own identifier and making statements about this resource (reification). We decided to follow the solution of MSC 2010 as closely as possible and used reification. The entity representing the statement to be commented on, is typed as an instance of \emph{mscvocab:SeeForStatement}, a subclass of \emph{rdf:Statement} newly introduced in 2020's \emph{mscvocab} extension. Differing from MSC 2010, we described a statement's \emph{rdf:subject}, \emph{rdf:predicate}, and \emph{rdf:object} (instead of MSC 2010's \emph{mscvocab:forTarget}). We provided the condition for the \emph{mscovab:seeConditionally} relation between two concepts via \emph{mscvocab:scope} just as in MSC 2010:

[fontsize=\small]
\begin{verbatim} 
msc:SeeForStatement-03B45-to-03B44 rdf:type owl:NamedIndividual ,
    mscvocab:SeeForStatement ;
    rdf:object <http://msc.org/resources/MSC/msc2020/03B44> ;
    rdf:predicate mscvocab:seeConditionally ;
    rdf:subject <http://msc.org/resources/MSC/msc2020/03B45> ;
    mscvocab:scope "for temporal logic"^^xsd:string .
\end{verbatim}

\begin{figure}
\centering
\includegraphics[width=.8\textwidth]{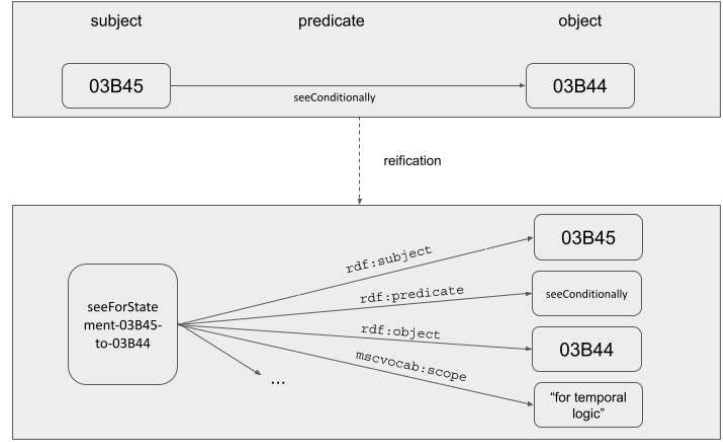}
\caption{The statement with predicate \emph{mscovab:seeConditionally} is characterised by a \emph{mscvocab:SeeForStatement} to account for the accurate scope.} \label{fig:seeCond}
\end{figure}

The idea behind \emph{mscovab:seeConditionally} is roughly illustrated in Figure~\ref{fig:seeCond}.

Subsequently, in a fourth step, we modeled the concept hierarchy (\emph{skos:broader / skos:narrower}). In contrast to the predecessor, we stated the hierarchical relation using \emph{skos:broader}, leaving \emph{skos:narrower} implied. This required assigning every lower-level concept to its appropriate super-ordinate concept. The information was acquired by further manipulating table data from the database excerpt with OpenRefine. 

Modeling the historical changelog from the MSC 2010 to 2020 was the fifth step. Instead of documenting the evolution of all four MSC generations in one model, we only considered the latest, consecutive two versions. This approach is in line with the idea of Linked Open Data, because it allows for semantic inference. To account for all information about MSC evolution in the first version made sense, however, because there was no prior version to refer to, i.~e., infer from.

Modeling decisions include: introduction of a new class in the MSC 2020, that did not exist in its predecessor, e.~g., 05-11, the documentation of an MSC 2010 class that has been removed, e.~g., 80M25, tracking changes within an MSC class, e.~g., 01-01, and making changes in a subordinate class  visible on the super-ordinate level, e.~g. for 05-XX because 05-11 has been newly added.

Since the evolution from MSC 2010 to 2020 is based on community discussions, we provided an intellectually compiled list of changes. Subsequently, the changes were modeled as in the MSC 2010's approach by using \emph{skos:mappingRelation} via OpenRefine and Prot{\'{e}}g{\'{e}}.

In a sixth step, we created collections. The MSC 2010 SKOS version already provided meta collections of historical works, proceedings, or computational methods across all classes. We stuck to this approach and introduced another collection for research data (all newly introduced -11 classes). 

The final seventh step concerned MSC specific scope notes. In contrast to a general \emph{skos:scopeNote} the MSC provides three scope notes, namely \emph{mscvocab:NotUseScopeNote}, \emph{mscvocab:MustUseScopeNote}, \emph{mscvocab:UseScopeNote}, with significant semantic differences. The prospective content was identified by the respective terms `do not use', `must', or `use' given in references in \emph{skos:prefLabel}'s parentheses. The process is the same for all three scope notes, but we consider \emph{mscvocab:UseScopeNote} as an example: Instances of \emph{mscvocab:UseScopeNote} were created, and triples between 
\begin{enumerate} 
    \item the \emph{skos:Concept} and the newly created instances of \emph{mscvocab:UseScopeNote} with \emph{skos:scopeNote} and
    \item the newly created instances of \emph{mscvocab:UseScopeNote} and the value of \emph{skos:prefLabel} with \emph{mscvocab:scope}
\end{enumerate} 
were derived.

The final public version of MSC 2020 is expected offer a number of alternate formats
for the collected data, as the MSC 2010 did on http://ms2010.org, such as its MediaWiki form,  various printable PDFs, KWIC indices, and even the TiddlyWiki tool (done at MR but not yet public).

\section{Conclusion \& Future Work}\label{sec.concl}
Our main objective was to provide a consistent, valid, complete MSC 2020 SKOS version for use-cases in knowledge organisation mainly motivated by uses in libraries. The SKOS version is largely similar from its predecessor but features some improvements with respect to the quality of the data itself and the modeling. It does mean that MSC information available as LOD will be up to date.

Of course, there are still short-term requirements:

As for the MSC 2010, an infrastructure for the MSC 2020 SKOS version needs to be provided: URIs must resolve correctly and meaningfully, an official website must be provided with the data itself, its documentation, and a SPARQL endpoint. This landing page should be linked to a development repository where the SKOS model can be further refined.  These are ongoing aspects fo the project which have not been finished yet.
As mentioned above, the editorial aspects of MSC 2020 have only just been finalized.

The small number of actual changes to MSC (on the order of hundreds) means that 
some of the pending additions planned, such as the descriptive text from other languages
will and relationships to DDC and UDC will carry over relatively simply from MSC 2010.  
For instance, not many new translations are needed.  They additions have not yet been made public as of this text's writing, but the project continues. 

In the future, we would also like to address the following desiderata:

Establish an editorial process that allows for supervised additions (e.~g., more languages or discussions) based on the SKOS model representation.

Provide a broad agreement on the modeling decisions and appropriate documentation to facilitate the transition from MSC 2020 to 2030.
\printbibliography

\end{document}